\documentclass[
    ,final            
  ]
  {aipproc}

\usepackage{amsmath}
\layoutstyle{6x9}

\def\He{$^3\!$~He}
\def\be{\begin{equation}}
\def\ee{\end{equation}}
\def\bea{\begin{eqnarray}}
\def\eea{\end{eqnarray}}
\def\jum#1{\hspace*{#1cm}}
\def\ju{\hspace*{0.5cm}}
\def\vum#1{\vspace*{#1cm}}
\def\mezzo{\frac{1}{2}}
\def\qu{q^+}
\def\qd{q^-}
\def\qz{q^0}
\def\GeVq{GeV~$\!^2$\ }

\def\Hermes{{\sc Hermes\ }}

\def\bd{b_1^d}
\def\g1d{g_1^d}

\def\Dg{D_{\gamma}}
\def\Np{n^+}
\def\Nm{n^-}
\def\Nz{n^0}
\def\Ap{A_{||}}
\def\At{A_T}

\def\su{\sigma_U}
\def\sp{\sigma^+}
\def\sm{\sigma^-}
\def\sz{\sigma^0}

\def\Ap{A_1}
\def\At{A_{zz}}

\def\qu{q^+}
\def\qd{q^-}
\def\qz{q^0}

\def\Journal#1#2#3#4{{#1} {\bf #2}, #3 (#4)}

\def\NIMA{{\em Nucl. Instrum. Methods} A}

\def\NPB{{\em Nucl. Phys.} B}
\def\PLB{{\em Phys. Lett.}  B}

\def\PRD{{\em Phys. Rev.} D}
\def\PRC{{\em Phys. Rev.} C}

\def\JPG{{\em J. Phys.} G}
\def\PAN{{\em Phys. At. Nucl.} G}
\def\PRP{{\em Phys. Rept.} }

\begin{document}

\title{
A First Measurement of the \\ Tensor-Polarized Structure Function $\bf \bd$}

\author{Marco Contalbrigo}{
  address={\vum{-0.4} (on behalf of the HERMES Collaboration) \\
\vum{0.3} INFN - Sezione di Ferrara e Dipartimento di Fisica dell'Universit\`a di Ferrara \\
Via del Paradiso 12, 44100~Ferrara, ITALIA}
}



\vum{-0.5}
\begin{abstract}
The \Hermes experiment studies the spin structure of the nucleon using the $27.6$ GeV longitudinally 
polarized positron beam of {\sc HERA} and an internal target of pure gases.  In addition to the well-known
spin structure function $g_1$, measured precisely with longitudinally polarized proton and deuteron 
targets, the use of a tensor-polarized deuteron target provides access to the tensor polarized 
structure function $\bd$. The latter, measured with an unpolarized beam, quantifies the dependence of
the parton momentum distribution on the nucleon spin.
\Hermes had a 1-month dedicated run with a tensor polarized deuterium target during the 
2000 data taking period. Here preliminary results on the tensor-polarized structure function $\bd$
are presented for the kinematic range $0.002<x<0.85$ and $0.1$~GeV~$\!^2$~$<Q^2<20$~GeV~$\!^2$.
\end{abstract}

\maketitle


\vum{-1.0}
\section{Introduction}
The \Hermes experiment \cite{hermes} has been designed to measure the nucleon spin structure 
functions from deep inelastic scattering (DIS) of polarized positrons and electrons from
polarized gaseous targets (H, D, \He). 
Three of the main leading-twist structure functions are listed in the table below, 
along with their interpretation in the Quark-Parton Model.
The sums are over quark and antiquark flavours $q$
and the dependences on $Q^2$ and Bjorken $x$ are omitted for simplicity:

\vum{0.2}
\begin{tabular}{cccc}
\vum{0.1} \jum{2.0} & Proton &\jum{1}& Deuteron \\
\vum{0.1} $ F_1$ & $ \mezzo \sum_q e_q^2 \,[\qu+\qd]$ && $ \frac{1}{3}\sum_q e_q^2 \,[\qu + \qd +\qz$]\\  
\vum{0.1} $ g_1$ & $ \mezzo\sum_q e_q^2 \,[\qu-\qd]$ && $ \mezzo\sum_q e_q^2 \,[\qu-\qd]$  \\
\vum{0.1} $ b_1$ & $--$ && $ \mezzo \sum_q e_q^2 \left[2\qz-(\qd+\qu)\right]$ \\
\end{tabular}
\vum{0.1}

The unpolarized structure function $F_1$ measures the quark momentum distribution summed
over all the possible helicity states.
The polarized structure function $g_1$ is sensitive to the spin structure of the 
nucleon, and measures the imbalance of quarks with the same ($q^+$) or opposite ($q^-$) helicity with 
respect to the nucleon they belong to. For targets of spin 1 such as the deuteron, the tensor 
structure function $b_1$ compares the quark momentum distribution between the zero-helicity state 
of the hadron ($q^0$) and the average of the helicity-1 states ($q^+ + q^-$).
As the deuteron is a weakly-bound state of 
spin-half nucleons, $\bd$ was initially predicted to be small~\cite{jaffe}.
More recently, coherent double scattering 
models have predicted a sizable $\bd$ at low $x$~\cite{nikolae,weise,bora}, violating the sum rule 
which suggests a vanishing first moment of $\bd$~\cite{close}. Although
$\bd$ describes basic properties of the spin-1 deuterium nucleus, and may affect the experimental
determination of $g_1^d$, it has not yet been measured. 
In 2000, \Hermes collected a dedicated data set with a tensor polarized deuterium target 
for the purpose of making a first measurement of $\bd$.
Preliminary results from these data are presented in this paper.

\vum{-0.5}
\section{Hermes setup}
The \Hermes experiment is installed in the HERA ring where the beam positrons self-polarize by emission
of synchrotron radiation (Sokolov-Ternov effect) along the direction of the bending magnetic field.
Longitudinal beam polarization, needed for the $\g1d$ measurement, is obtained with two spin rotators
placed immediately upstream and downstream of the \Hermes apparatus.
As $\bd$ appears in the symmetric part of the hadronic tensor~\cite{jaffe}, 
its measurement neither depends on the beam polarization nor is diluted by the depolarization 
of the virtual photon. At \Hermes, an ``unpolarized beam'' is achieved by grouping together data
with opposite beam polarization.

A unique feature of the \Hermes experiment is its tensor-polarizable gaseous target~\cite{target}. 
Deuterium has a total of six hyperfine states depending on the relative orientation of the electron 
and nuclear spins. An atomic beam source (ABS) generates a Deuterium atomic beam and selects the 
two hyperfine states with the desired nuclear 
polarization. A cylindrical 40 cm long, $75\,\mu m$ thick Al tube (target cell) confines the polarized 
gas along the positron beam line, where a longitudinal magnetic field provides the quantization axis 
for the nuclear spin. Every $90$~seconds, a diagnostic system measures the atomic 
and molecular abundances 
and the atomic polarization inside the cell, then the polarization of the injected gas is changed.
The vector $V$ and tensor $T$ atomic polarizations are 
\bea
V\,=\,\frac{\Np-\Nm}{\Np+\Nm+\Nz} &\jum{2}& T\,=\,\frac{\Np+\Nm \,-\, 2\Nz}{\Np+\Nm+\Nz},
\eea
where $\Np,\,\Nm,\,\Nz$ are the atomic populations inside the cell with positive, negative and zero 
spin projection onto the beam axis. In 2000 the running conditions were very stable, 
the recombination on the cell surface was negligible, 
and no correction was needed for the residual polarization of the recombined atoms.
An average tensor polarization greater than $80$~\%  was obtained 
with a residual vector polarization of only $1$~\%. In contrast to the
solid targets used by other experiments, the \Hermes target does not suffer from radiation
damage thanks to the continuous flow of the target gas. 
Its polarization can be continuously measured, rapidly reversed,
and is not diluted by non-polarizable material. 

The \Hermes detector is a forward spectrometer with a dipole magnet providing a field integral of
1.3 T$\cdot$m. A horizontal iron plate shields the HERA beam lines from the field, thus dividing the 
spectrometer into two identical halves with a minimum vertical acceptance of $\pm 40$ mrad. The 
acceptance extends to $\pm 140$ mrad vertically and to $\pm 170$~mrad horizontally. 
In this analysis $36$ drift chamber planes in each detector half were used for tracking.
Positron identification is accomplished using a probability method based on signals of 
three subsystems: a lead-glass block calorimeter, a transition-radiation detector, and a preshower 
hodoscope. For positrons in the momentum range of $2.5$ to $27$ GeV, the identification efficiency 
exceeds 98~\% with a negligible hadron contamination, the average polar angle resolution is $0.6$ 
mrad, and the average momentum resolution is 1~--~2~\%.



\vum{-0.4}
\section{Measurement}
Depending on the beam ($P_B$) and target ($V$ and $T$) polarizations, the lepton-nucleon DIS cross 
section measured by the experiment is sensitive to the vector $\Ap$ ($A_2$ is here neglected) 
and tensor $\At$ asymmetries of the virtual photon nucleon cross section 
%

\vum{-0.6}
\bea
\sigma_{\rm meas} \,=\, \su \left[1 \,+\, P_B \Dg V \Ap \,+\, \mezzo T \At\right] &\jum{3}
\begin{matrix}
 & P_B & V & T \\
 \sp & 1 &  1 & 1 \\
 \sm &  1 &  -1 & 1 \\
  \sz & 0 & 0 & -2 \\
\end{matrix} \jum{1}
\eea
$\su$ is the unpolarized cross section $(\sp + \sm + \sz) / 3$ and
$\Dg$ is polarization transfer from the lepton beam to the virtual photon. Here and 
in the following no higher-twist contribution ($g_2,\,b_3,\,b_4$) is considered. 
The vector asymmetry $\Ap$ is related to $\g1d$; it is measured with a polarized beam 
($P_B = 1$) by comparing the cross sections for a target with the same $\sp$ and opposite $\sm$ 
helicity as the beam:

\vum{-0.4}
\bea
\Ap &=& \frac{\sp \,-\, \sm}{ 2\su\Dg} 
\ \ \ = \ \ \ \frac{g_1}{F_1} 
\ \ \ = \ \ \ A_1^\mathrm{meas}\ \left[1+\mezzo T\At\right]  
\;.\eea
Here $A_1^\mathrm{meas}$ is the asymmetry typically measured by experiments,
where $(\sp + \sm)/2$ is used rather than $\su$. 
The tensor term, until now neglected in $\g1d$ measurements, is needed to ensure that the denominator 
is proportional to $\su$ and so to $F_1$. The tensor asymmetry $\At$, and the related $\bd$, 
provide the missing
information on the difference between the cross section 
for the zero-helicity target state and the spin-averaged states of helicity 1:

\vum{-0.6}
\bea
\At &=& \frac{(\sp+\sm) \,-\, 2\sz}{3\su} 
\ju = \ju - \frac{2}{3} \frac{b_1}{F_1}
\;.\eea
At \Hermes, the target residual vector polarization in the $\sz$ case and in the case of the sum 
$\sp +\sm$ is very 
small and has a negligible effect on the $\At$ measurement. The vector contribution is further reduced
by combining together data with opposite beam helicities.
The kinematic range accessible for the $\bd$ measurement is limited to the intervals
$0.002 < x < 0.85$ and $0.1 < y < 0.91$ due to detector acceptance, resolution, and trigger requirements.
Inclusive deep-inelastic events
are selected by the requirements: $Q^2>0.1$ \GeVq and $W^2>3.24$ \GeVq. 
The total $x$-range is divided into 6 bins. 
For each $x$-bin the following yield asymmetry is calculated:

\be
\At \,=\,
\frac{1}{T} \cdot \left[ \frac{(N/L)^+ + (N/L)^- - 2 \cdot (N/L)^0}{(N/L)^+ + (N/L)^- + (N/L)^0} \right]
\;.\ee
The numbers of events selected ($N$)
are corrected per spin state for the background arising from charge symmetric processes. 
The corresponding luminosities ($L$) used for normalization are measured from Bhabha 
scattering off the target gas electrons. 
The radiative corrections on $\At$ are calculated using POLRAD \cite{polrad}. For this preliminary 
result the DIS radiative tail is neglected due to the small size of $\At$ and the 
elastic~radiative~tail
\begin{figure}[th]
\parbox{0.54\linewidth}
{
\includegraphics[width=1.\linewidth]{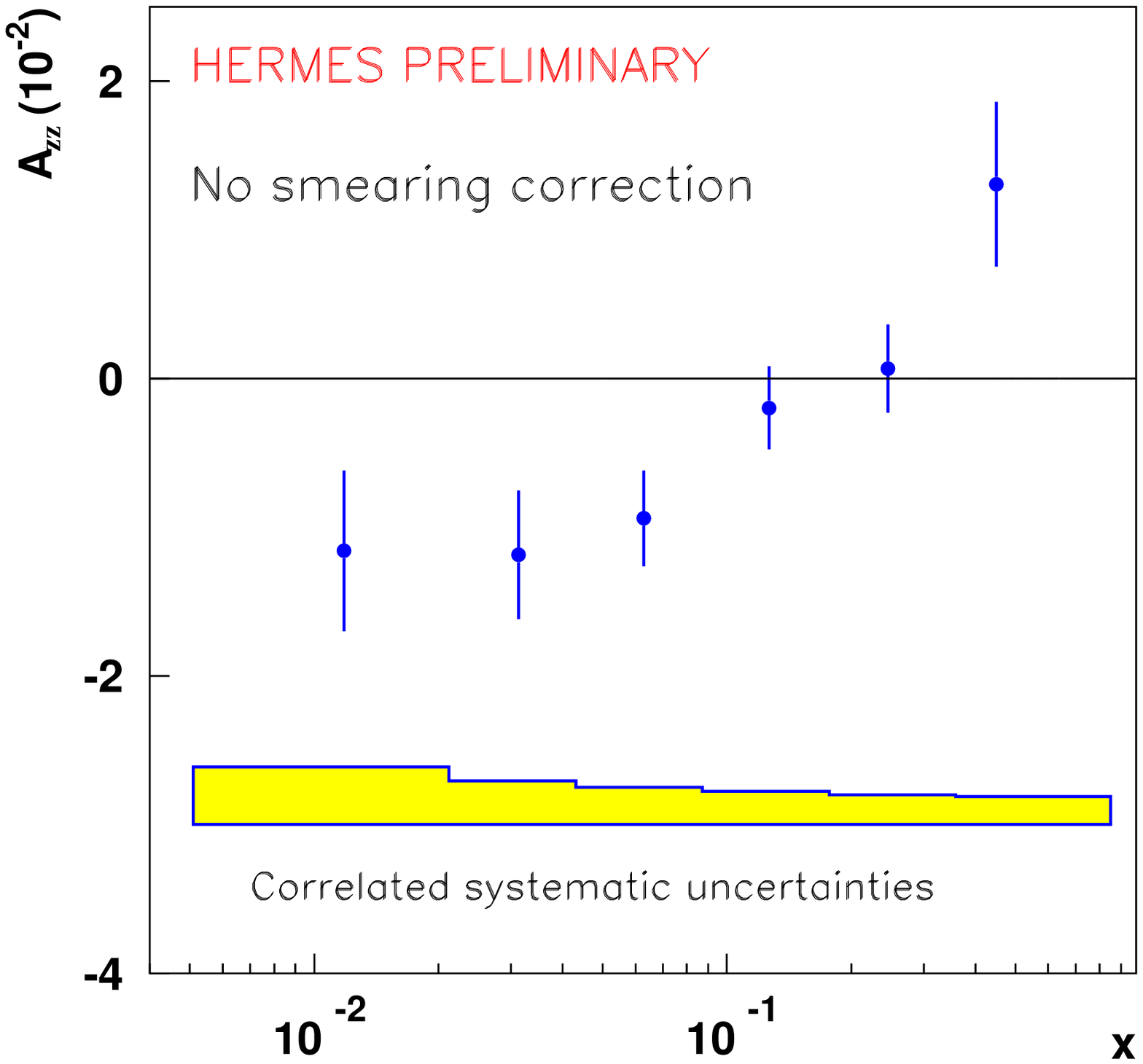}
}
\parbox{0.44\linewidth}
{
\includegraphics[width=1.\linewidth]{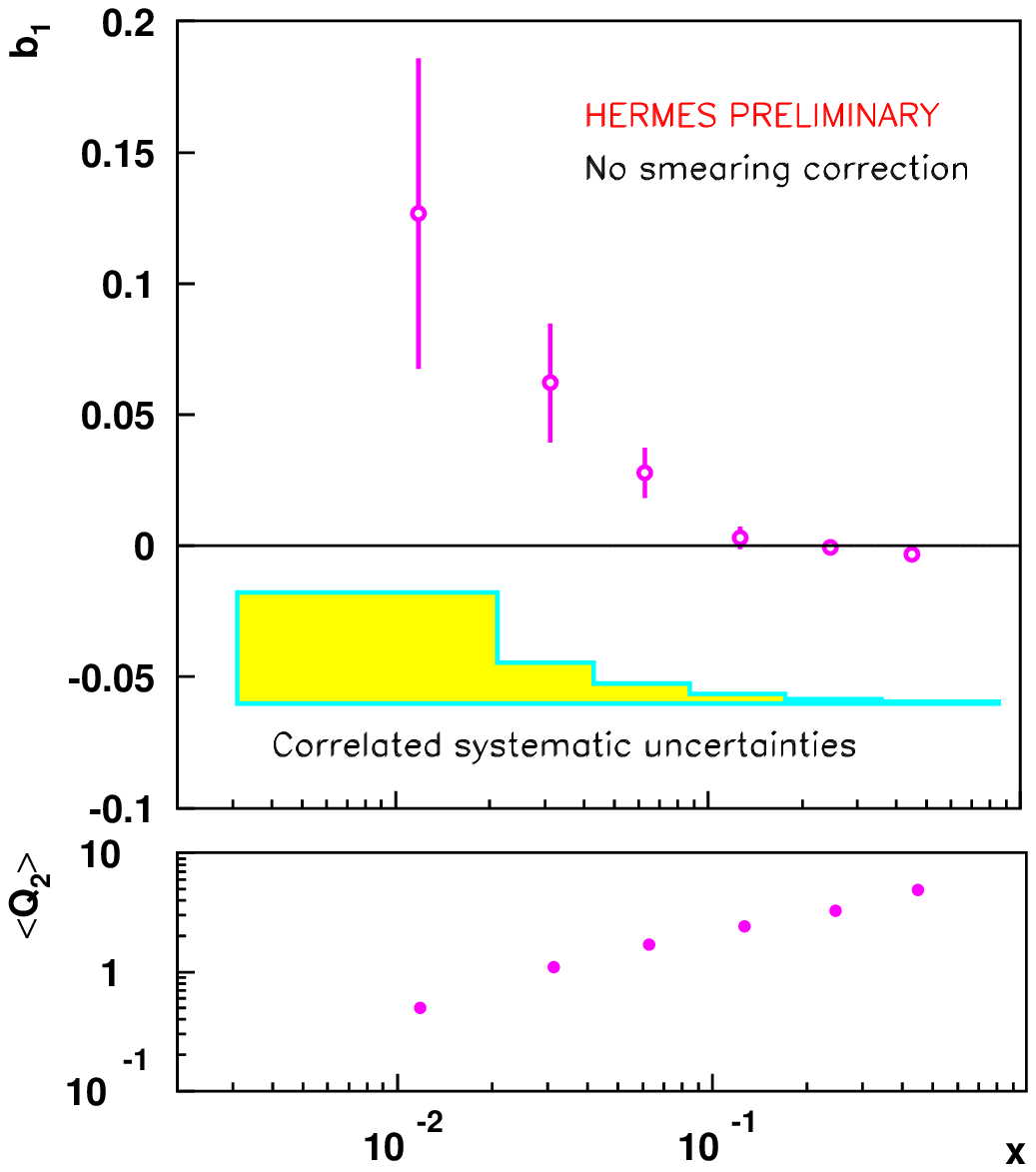}
}
\caption{The tensor asymmetry $\At$ (left) and the tensor-polarized structure function $\bd$ (right) 
as measured by \Hermes. The error bars are statistical only and the shaded bands show the estimated 
systematic uncertainties. The bottom panel on the right plot shows the average $Q^2$ of the 
measurements.}
\label{Figuno}
\end{figure}
is estimated from a not fully updated parameterization of the deuteron quadrupole form 
factor~\cite{quadru}.  The statistical error 
was enlarged by almost a factor of 2 at low $x$ to account for radiative background.
The measured $\At$ is presented in Fig. \ref{Figuno}. The systematic uncertainties are correlated 
over the kinematical bins, being dominated by the target density normalization between different 
injection modes of the ABS. The target polarization measurement, the misalignment of the spectrometer,
and the hadron contamination give negligible systematic effects. Data with different beam 
polarizations were checked to give compatible $\At$ results. No systematic error has been estimated from the 
still incomplete radiative correction. $\At$ is found to be less than~$2$~\%. From this result, the 
influence of the tensor asymmetry bias on a $\g1d$ measurement is estimated to be less than 0.5~--~1.0~\%.

\begin{center}
\begin{figure}[h]
\includegraphics[width=0.48\linewidth]{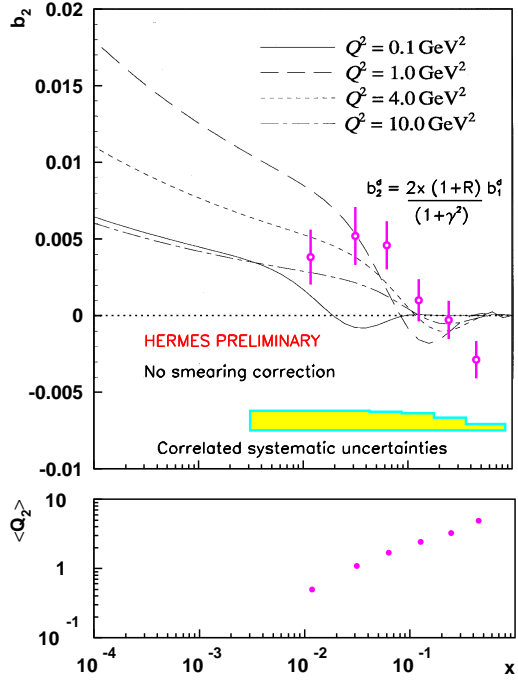}
\caption{The tensor-polarized structure function $b_2^d$. 
The error bars are statistical only and the shaded bands show 
the estimated systematic uncertainties.
The bottom panel shows the average $Q^2$ of the measurements. The curves are from calculations
within the re-scattering model of Ref.~\cite{bora} for $Q^2$ values in the range of the \Hermes
measurement.}
\label{Figdue}
\end{figure}
\end{center}

The spin structure function $\bd$ is extracted from the tensor asymmetry via the relation
%
$\bd\,=\, -\frac{3}{2}\At \, \frac{(1+\gamma^2) F_2^d }{2x(1+R)} $, 
where the structure function $F_1^d$ has been expressed in terms of the ratio
$R=\sigma_L/\sigma_T$~\cite{ssratio} and the structure function $F_2^d$ ($\gamma$ is a 
kinematic factor).  $F_2^d\,=\,F_2^p (1+F_2^n/F_2^p)/2$ is calculated using 
parameterizations for $F_2^p$~\cite{allm97} and $F_2^n/F_2^p$~\cite{ffratio}.
Fig.~\ref{Figuno} displays the result for $\bd$, which is small but different from zero.
The structure function $b_2^d$ has also been extracted, using the Callan-Gross relation
$
b_2^d \,=\, \frac{2x(1+R)}{(1+\gamma^2)} \bd
$.
The data indicate a rise at low $x$ as predicted by the most recent theoretical 
models~\cite{nikolae,weise,bora}. A comparison of the measured $b_2^d$ with the prediction of one of the 
models~\cite{bora} is given in Fig.~\ref{Figdue}. 

In conclusion \Hermes has provided the first direct measurement of the 
structure function $\bd$ in the kinematic range $0.002<x<0.85$ and $0.1$~GeV~$\!^2$~$<Q^2<20$~GeV~$\!^2$.~The
preliminary result for the tensor asymmetry is sufficiently small to produce an effect of 
%
less than 1~\% on the measurement of $\g1d$.
The dependence of $\bd$ on Bjorken $x$ is in qualitative agreement with expectations 
based on coherent double scattering models~\cite{nikolae,weise,bora} and favors a sizeable $\bd$ at 
low-$x$. This suggests a significant tensor 
polarization of the sea-quarks, violating the Close-Kumano sum rule~\cite{close}. This observation
is analogous to the well-established $\overline{u}-\overline{d}$ asymmetry of the sea 
and the consequent violation of the Gottfried sum rule~\cite{gott}.




\vum{-0.5}



\begin{thebibliography}{99}

\vum{-0.3}
\bibitem{hermes} \Hermes Coll., K. Ackerstaff {\it et al}, \Journal{\NIMA}{417}{230}{1998}
\bibitem{jaffe} P. Hoodbhoy {\it et al}, \Journal{\NPB}{312}{571}{1989},
                H. Khan {\it et al}, \Journal{\PRC}{44}{1219}{1991}
\bibitem{nikolae} N. N. Nikolaev {\it et al}, \Journal{\PLB}{398}{245}{1997} 
\bibitem{weise} J. Edelmann {\it et al}, \Journal{\PRC}{57}{254}{1998} 
\bibitem{bora} K. Bora {\it et al}, \Journal{\PRD}{57}{6906}{1998}
\bibitem{close} F. E. Close {\it et al}, \Journal{\PRD}{42}{2377}{1990}
\bibitem{target} C. Baumgarten {\it et al}, \Journal{\NIMA}{482}{606}{2002}
\bibitem{polrad} I. V. Akushevich {\it et al}, \Journal{\JPG}{20}{513}{1994}
\bibitem{quadru} Kobushkin {\it et al}, \Journal{\PAN}{58}{1477}{1995}
\bibitem{ssratio} L. W. Whitlow {\it et al}, \Journal{\PLB}{250}{193}{1990}
\bibitem{allm97} ALLM97 parameterization: H. Abramowicz {\it et al}, hep-ph 9712415
\bibitem{ffratio} NMC Coll., P. Amaudruz {\it et al}, \Journal{\NPB}{371}{3}{1992}
\bibitem{gott} S. Kumano, \Journal{\PRP}{303}{183}{1998}
\end{thebibliography}
\end{document}